\documentclass[conference]{IEEEtran}
\IEEEoverridecommandlockouts

\newcommand\hmmax{0}
\newcommand\bmmax{0}

\usepackage[ruled,vlined,linesnumbered]{algorithm2e} 
\usepackage{graphicx}
\usepackage{tikz}
\usepackage{pgfplots}
\usepackage{xcolor}
\usepackage{color, colortbl}
\usepackage{amsmath}
\usepackage{bbm}
\usepackage{amsfonts, amssymb}
\usepackage{mathtools}
\usepackage{enumerate}
\usepackage{cite}
\usepackage[nolist]{acronym}
\usepackage{amsmath}

\usepackage{booktabs} 		
\usepackage{diagbox}		
\usepackage{upgreek}
\usepackage{bbm}
\usepackage{Files/bm}
\usepackage{Files/winsnotation}
\usepackage{Files/Symbols_OPL}
\usepackage{Files/LVcolours} 

\usepackage{setspace}	 	
\usepackage{comment}
\usepackage{physics}
\usepackage{quantikz}
\usepackage{enumitem}

\newlist{myitemize}{itemize}{3}
\setlist[myitemize,1]{label=1.,leftmargin=1em}
\setlist[myitemize,2]{label=$\rightarrow$,leftmargin=0.75em}
\setlist[myitemize,3]{label=$\diamond$}

\usepackage{array}
\newcolumntype{C}[1]{>{\centering\arraybackslash}p{#1}}

\makeatletter
\def\endthebibliography{%
  \def\@noitemerr{\@latex@warning{Empty `thebibliography' environment}}%
  \endlist
}
\makeatother
\usepackage{amsthm}
\theoremstyle{definition}

\newtheorem{example}{Example}
\usepackage{algpseudocode,amsmath}

\usetikzlibrary{arrows}
\usepgfplotslibrary{fillbetween}

\usetikzlibrary{automata, positioning}

\pgfplotsset{compat=1.17}

\begin{document}

\title{Dimensioning of Quantum Memories for \\ Distilled Quantum EPR Packets}
\author{%
\IEEEauthorblockN{Lorenzo~Valentini, Diego~Forlivesi, Andrea~Talarico, Marco~Chiani}
\IEEEauthorblockA{CNIT/WiLab, DEI, University of Bologna, Italy \\
Email: \{lorenzo.valentini13, diego.forlivesi2, andrea.talarico3,  marco.chiani\}@unibo.it }
}


\maketitle 
\markboth{}{}

\begin{acronym}
\small
\acro{AWGN}{additive white Gaussian noise}
\acro{BC}{bubble clustering}
\acro{BCH}{Bose–Chaudhuri–Hocquenghem}
\acro{CDF}{cumulative distribution function}
\acro{CPMG}{Carr-Purcell-Meiboom-Gill}
\acro{CRC}{cyclic redundancy code}
\acro{DEJMPS}{Deutsch--Ekert--Jozsa--Macchiavello--Popescu--Sanpera}
\acro{EPR}{Einstein-Podolsky-Rosen}
\acro{LDPC}{low-density parity-check}
\acro{LUT}{lookup table}
\acro{ML}{maximum likelihood}
\acro{MWPM}{minimum weight perfect matching}
\acro{QECC}{quantum error correcting code}
\acro{PDF}{probability density function}
\acro{PMF}{probability mass function}
\acro{MPS}{matrix product state}
\acro{WEP}{weight enumerator polynomial}
\acro{WE}{weight enumerator}
\acro{BD}{bounded distance}
\acro{QLDPC}{quantum low density parity check}
\acro{CSS}{Calderbank, Shor, and Steane}
\acro{MST}{minimum spanning tree}
\acro{PruST}{pruned spanning tree}
\acro{RFire}{Rapid-Fire}
\acro{UF}{union-find}
\acro{LEMON}{library for efficient modeling and optimization in networks}
\acro{STM}{spanning tree matching}
\acro{i.i.d.}{independent identically distributed}
\acro{SC}{sphere clustering}
\acro{QEC}{quantum error correction}

\end{acronym}
\setcounter{page}{1}

\begin{abstract}
The quantum Internet envisions a network where information is transmitted through entanglement, with Einstein–Podolsky–Rosen (EPR) pairs serving as one of the fundamental carriers. 
In this work, we propose a framework for dimensioning quantum memories capable of storing distilled EPR pairs useful to transmitting and manage quantum error correcting codes. 
Using a Markov chain model, we capture the stochastic evolution of stored entangled states in quantum memories, linking memory performance to system parameters such as technology characteristics and initial entanglement fidelity. 
Building on this framework, we provide analytical tools and design principles for optimizing memory architectures that preserve high-fidelity entanglement over time, ensuring the availability of encoded quantum resources necessary for several operations in future quantum Internet infrastructures transmitting EPR packets.
\end{abstract}


\section{Introduction}


As communication technology progresses beyond classical paradigms, the quantum Internet has emerged as a transformative vision for future information networks~\cite{Mun2015:QRepeater,WehElkHan18:QInternet,Cac19:QInternet,Pom22:experimentalQI}. 
Unlike the classical Internet, which relies on bits, a quantum Internet enables the transmission and manipulation of qubits, quantum states that can exist in coherent superpositions. 
At the heart of this quantum communication infrastructure lies one of the most fundamental resources in quantum mechanics: entanglement~\cite{rfc9340}. 
In particular, \ac{EPR} pairs are envisioned to form the basic units of quantum connectivity. Through entanglement, quantum information can be reliably transferred via quantum teleportation~\cite{Ben93:teleporting}, establishing a fundamental communication mechanism for the quantum Internet.


Beyond quantum communication, \ac{EPR} pairs are central to scalable quantum computing architectures. In modular processors, where qubits are organized into subsystems or logical units protected by quantum error-correcting codes, \ac{EPR} pairs provide entanglement links between modules~\cite{Yod25:IBM_tourgross}. These links enable coherent operations across logical codewords without direct physical coupling, effectively extending computation beyond a single module~\cite{Yoder25:EPRdiversiModuli}. In distributed quantum computing, \ac{EPR}-based entanglement allows physically separated processors to operate as a unified system, forming a quantum data center~\cite{Cisco25:DataCenter}. In this setting, \ac{EPR} pairs support logical-qubit teleportation and nonlocal gate execution, offering a scalable framework for resource sharing and parallelism.

Quantum memories are essential components of quantum information engineering, enabling the storage and retrieval of quantum states while preserving coherence~\cite{Ter:15}. 
They are therefore critical to both quantum communication and computing. 
In~\cite{els24:fidelity}, a protocol employing two long-term and one short-term memory was proposed to maintain high-fidelity entangled pairs. 
In~\cite{dav24:entanglement}, availability and average fidelity were studied for a two-qubit memory, where each generated \ac{EPR} pair was either distilled to improve fidelity or discarded upon failure, illustrating the trade-off between fidelity and availability. 
In contrast to these works, we investigate quantum memory dimensioning for transmitting and manipulating quantum states encoded across multiple physical qubits, as required in quantum error correction-based systems~\cite{BraKit:98, ForValChi24:JSAC, Bra24:BBCodes, ValForChi25:CylMob}.

In our vision of quantum Internet, communication between two nodes is not limited to the on-demand generation and distribution of a single entangled pair for each quantum transaction. 
Instead, we imagine that entire \emph{\ac{EPR} packets} are shared between nodes that wish to establish a quantum link, analogous to how packets of bits are exchanged in classical networks, rather than individual bits being transmitted when required. 
Each \ac{EPR} packet would contain multiple entangled pairs, collectively serving as a reservoir of quantum correlations that can be consumed by higher-level communication protocols or enhanced through distillation in case of high reliability requirements. 
In this framework, the \ac{EPR} packet becomes the quantum counterpart of the classical data packet, an encapsulated unit of quantum resources that can be routed, stored, and consumed dynamically as part of the broader quantum network infrastructure.

In this paper, we address the dimensioning of quantum memories for storing entangled qubits between devices and provide a detailed analysis of the evolution of the memory states over time.
The primary contributions of this work lie in advancing the modeling, design, and operational protocols of quantum memories for distilled \ac{EPR} pairs. 
First, we develop a framework to describe the fidelities of the \ac{EPR} pairs stored in quantum memories, formulating the problem in terms of a Markov chain to capture the stochastic evolution of entangled states over time, considering that a portion of them are consumed at each round. 
Building on this framework, we introduce mathematical tools that enable the systematic design of quantum memories, linking their performance to key system parameters such as the underlying technology characteristics and the initial fidelity of entangled states as determined by the entanglement distribution process. 
Finally, we propose a protocol to reliably generate and manage multiple \ac{EPR} pairs, ensuring that sufficient high-fidelity entanglement is available to operate quantum error correcting codes effectively. 


\section{Preliminaries and Background}
\label{sec:preliminary}

A qubit is an element of the two-dimensional Hilbert space $\mathcal{H}^{2}$, with orthonormal basis $\ket{0}$ and $\ket{1}$. 
The Pauli operators $\M{I}, \M{X}, \M{Z}$, and $\M{Y}$, are defined by  $\M{I}\ket{a}=\ket{a}$, $\M{X}\ket{a}=\ket{a\oplus 1}$, $\M{Z}\ket{a}=(-1)^a\ket{a}$, and $\M{Y}\ket{a}=i(-1)^a\ket{a\oplus 1}$ for $a \in \lbrace0,1\rbrace$ where $\oplus$ is the XOR operation. These operators either commute or anticommute with each other. 
We use the notation $\ket{\Phi^{\pm}} = (\ket{00} \pm \ket{11})/\sqrt{2}$ and $\ket{\Psi^{\pm}} = (\ket{01} \pm \ket{10})/\sqrt{2}$ for two-qubit Bell's states or, equivalently, \ac{EPR} pairs. 
We define a mixed state as a distribution over quantum states, $\{ p_i, \ket{\psi_i} \}$, meaning that with probability $p_i$ the system is in state $\ket{\psi_i}$. We represent the state of the quantum system using the density matrix representation $\M{\rho} = \sum_i p_i \ket{\psi_i} \bra{\psi_i}$.

\subsection{Quantum Distillation}
\label{subsec:QPurification}
Imperfect entanglement generation induces errors on the shared \ac{EPR} pairs, leading to mixed states. 
We describe the mixed state with the density matrix
\begin{align}
\label{eq:GenMixedState}
\M{\rho} &= A_0 \ket{\Phi^+}\bra{\Phi^+} + B_0 \ket{\Psi^-}\bra{\Psi^-} \notag \\
&+ C_0 \ket{\Psi^+}\bra{\Psi^+} + D_0 \ket{\Phi^-}\bra{\Phi^-}
\end{align}
where the coefficients are real, normalized $A_0 + B_0 + C_0 + D_0 = 1$, and defined on the interval $[0, 1]$. 
In this case, the coefficient $A_0$ is the entangled pair fidelity since we are considering teleportation using the $\ket{\Phi^+}$ state.
An important mixed state is the Werner state \cite{Wer89:WernerState, Ben96:purification, Zha02:WernerPrep}, having density matrix
\begin{align}
\label{eq:Werner}
\M{\rho} &= F_0 \ket{\Phi^+}\bra{\Phi^+} \notag \\
&+ \frac{1-F_0}{3} \Big[ \ket{\Psi^-}\bra{\Psi^-} + \ket{\Psi^+}\bra{\Psi^+} + \ket{\Phi^-}\bra{\Phi^-}\Big]
\end{align}
where $F_0$ is the fidelity of the raw \ac{EPR} pairs provided by the entanglement generation system.

Several techniques have been developed to increase the fidelity of entangled states for teleportation. 
In the following, we consider the \ac{DEJMPS} distillation algorithm presented in \cite{Deu96:purification} which improves \cite{Ben96:purification}.
Between symmetric distillation steps the state evolves as follow~\cite{Deu96:purification}
\begin{equation}
\label{eq:DeautschProbEvol}
\begin{aligned}
A_{i+1} &= (A_{i}^2 + B_i^2) \, N_i^{-1}\\
B_{i+1} &= 2 C_{i} D_i N_i^{-1}\\
C_{i+1} &= (C_{i}^2 + D_i^2) \, N_i^{-1}\\
D_{i+1} &= 2 A_{i} B_i N_i^{-1}\\
\end{aligned}
\end{equation}
where $N_i = (A_i + B_i)^2 + (C_i + D_i)^2$ and $i$ indicates the iteration step.
An important consideration is that distillation can fail, depending on the probabilities associated with the mixed-state components $A_i$, $B_i$, $C_i$, and $D_i$.
In particular, the probability that a state in the mixture $(A_i, B_i, C_i, D_i)$ is successfully distilled into the state $(A_{i+1}, B_{i+1}, C_{i+1}, D_{i+1})$~is
\begin{align}
    p_{i \to i+1} = N_i =  (A_i + B_i)^2 + (C_i + D_i)^2\,.
\end{align}
Due to this inherent randomness, the entangled qubits must be stored in quantum memories until both the transmitter and receiver confirm the success of the operation.

\section{Quantum Memory Dimensioning}
\label{sec:mainContrib}

\subsection{Memory Management Operations}
\begin{figure}[t]
 	\centering
 	\includegraphics[width=\columnwidth]{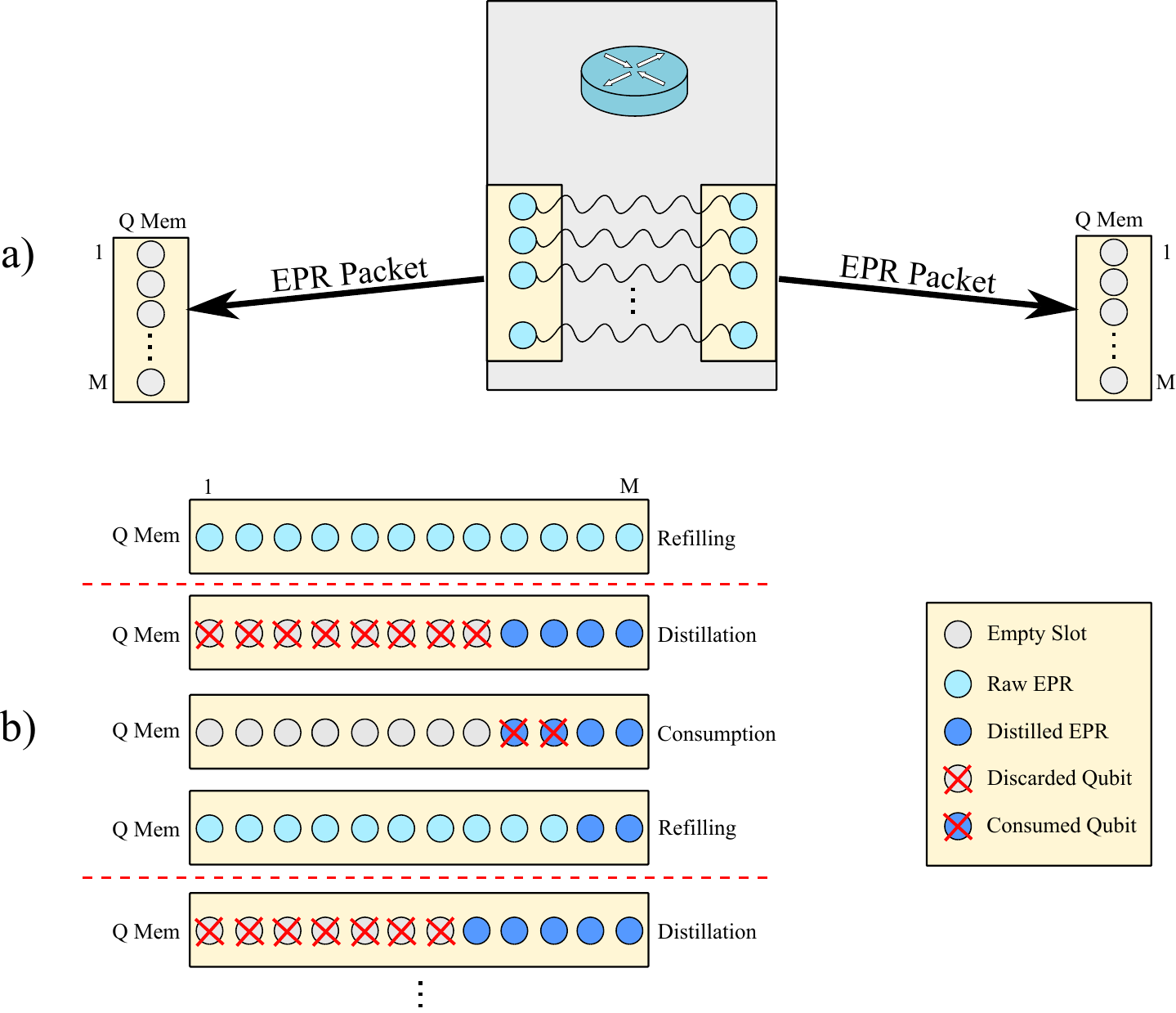}
 	\caption{System model for quantum memory dimensioning. (a) Entanglement distribution of EPR packets. (b) The three operations performed by the memory: distillation, consumption, and refilling.  }
 	\label{fig:qMemDes}
\end{figure}

We assume that between points $A$ and $B$ we have access to \ac{EPR} packets containing raw entangled qubits with initial fidelity $F_0$, modeled as Werner states.
This is pictorially shown in Fig.~\ref{fig:qMemDes}a. 
These qubits can be stored in two quantum memories, each capable of holding up to $M$ qubits (i.e., up to $M$ entangled pairs in total).  
Let $n_i$ denote the number of entangled pairs with fidelity $F_i$, where the index $i$ labels the distillation step.
Initially, the quantum memory is filled with $M$ entangled pairs, each having fidelity $F_0$, i.e., $n_0 = M$.
We denote by $d$ the maximum number of distillation steps supported by the system.  

Time is divided into rounds, and in each round the system performs three operations: $i)$ distillation; $ii)$ consumption; $iii)$ refilling.  
These three phases are exemplified in Fig.~\ref{fig:qMemDes}b.
During the distillation phase, \ac{DEJMPS} distillation is applied to the \ac{EPR} memory. 
In particular, we operate on $n_i$ pairs of fidelity $F_i$, producing at most $\lfloor n_i / 2 \rfloor$ new pairs of fidelity $F_{i+1}$, for $i = 0, \dots, d-1$.  
Note that, all these distillations can be performed in parallel.
During the consumption phase, if available, $c$ pairs at fidelity $F_d$ are consumed.  
During the refilling phase, the memory is replenished with new pairs of fidelity $F_0$, occupying the slots of consumed or discarded pairs. In this way, the memory is fully occupied at the start of each round, ensuring that  
\begin{equation}
    \sum_{i=0}^d n_i = M\,.
\end{equation}

\begin{example}
    Let us consider a quantum memory with $M = 16$, $d=2$, and $(n_0, n_1, n_2) = (4, 7, 5)$. Starting from the seven entangled qubits at $F_1$ we can obtain at most three entangled qubits at $F_2$ combining six of them. Assume this is the case. 
    Then, from the four entangled qubits at $F_0$ we can obtain at most two entangled qubits at $F_1$ combining all of them. Assume only one distillation has success.
    After this distillation phase we end up in a state with $(n_0, n_1, n_2) = (0, 1, 5) + (0, 1, 3) = (0, 2, 8)$, where $(0, 1, 3)$ are the new distilled qubits and $(0, 1, 5)$ are the qubits that are not distilled in this phase.
    Then, considering a consumption of one qubit, i.e., $c=1$, we obtain $(n_0, n_1, n_2) = (0, 2, 7)$. 
    Finally, we refill the memory with new raw entangled qubits at $F_0$, untill we occupy all memory slots, leading to $(n_0, n_1, n_2) = (7, 2, 7)$.
\end{example}


\subsection{Markov Chain Description}
\label{subsec:MarkovChain}

To model this procedure we use a Markov chain where $(n_0, n_1, \dots, n_d)$ represent a state $\ell$ in the finite state space $\mathcal{S}$.
The transition probabilities between these states are found computing the probability to have $(\tilde{n}_1, \dots, \tilde{n}_d)$ new distilled pairs, which is a product of binomial distributions as
\begin{align}
    p(\tilde{n}_1, \dots, \tilde{n}_d) = \prod_{i=0}^{d-1} \mathcal{B}(\lfloor n_i/2 \rfloor, \tilde{n}_{i+1}, p_{i \to i+1}) 
\end{align}
where $\mathcal{B}(n, k, p) = \binom{n}{k} p^k (1-p)^{n-k}$.
Then, considering consumption phase and refilling phase we observe that the probability $p(\tilde{n}_1, \dots, \tilde{n}_d)$ maps the state $(n_0, n_1, \dots, n_d)$ to the state $(n^\prime_0, n^\prime_1, \dots, n^\prime_d)$ where
\begin{align}
    n^\prime_i = \begin{dcases}
        \tilde{n}_i + (n_i \mod 2) & 0 < i < d \\
        \max\{\tilde{n}_d + n_d - c, 0\}& i = d \\
    \end{dcases} 
\end{align}
and $n^\prime_0$ is chosen such that $\sum_{i=0}^d n^\prime_i = M$ due to the refilling phase.
We denote as $\M{P}$ the transition probability matrix, where each entry specifies the probability of transitioning between states. 
In this way, if $\ell \in \mathcal{S}$ is the state represented by $(n_0, n_1, \dots, n_d)$ and $\ell^\prime \in \mathcal{S}$ is the state represented by $(n^\prime_0, n^\prime_1, \dots, n^\prime_d)$, then $P_{\ell,\ell^\prime} = p(\tilde{n}_1, \dots, \tilde{n}_d)$.

\begin{example}
    Let us consider a quantum memory with $M = 16$, $d = 2$, $c = 1$, $p_{0 \to 1} = 0.88$, and $p_{1 \to 2} = 0.87$. Suppose the system is in the state $(n_0, n_1, n_2) = (4, 7, 5)$. The probability of observing $(\tilde{n}_1, \tilde{n}_2) = (1, 3)$ is given by $p(\tilde{n}_1 = 1, \tilde{n}_2 = 3) = 2p_{0 \to 1}(1-p_{0 \to 1})p_{1 \to 2}^3 \approx 0.14$. This probability corresponds to a transition from the state $(n_0, n_1, n_2) = (4, 7, 5)$ to the state $(n_0, n_1, n_2) = (7, 2, 7)$.
\end{example}

\subsection{Asymptotic Analysis}
In this section we propose an asymptotic analysis that can be carried out within the proposed framework to design the system parameters.
Let consider a Markov chain with state space $\mathcal{S}$.  
Suppose that for some $\ell \in \mathcal{S}$, every state $i \in \mathcal{S}$ can reach $\ell$.
Then, the state $\ell$ belongs to a closed communicating class.
In fact, defining $\mathcal{C} = \{ k \in \mathcal{S} \mid \ell \leftrightarrow k \}$, where $\ell \leftrightarrow k$ means that $\ell$ and $k$ communicate (i.e., we can reach $k$ from $\ell$, $\ell \to k$, and vice versa). 
By construction, $\mathcal{C}$ is a communicating class. 
To verify that $\mathcal{C}$ is closed, suppose there exists $k \in \mathcal{C}$ and $m \in \mathcal{S}$ such that $k \to m$. If $m \notin \mathcal{C}$, then by definition $\ell$ cannot reach $m$. 
But since $\ell \to k$ and $k \to m$, the path from $m$ to $\ell$ would contradict the assumption that $m \notin \mathcal{C}$. Therefore, every state reachable from $k \in \mathcal{C}$ must also lie in $\mathcal{C}$. Hence, $\mathcal{C}$ is closed.

We denote by $\V{v}_i$ the distribution vector of the Markov chain at the step $i \ge 0$. 
The product $\V{v}\M{P}$ then yields a probability distribution over the states after one transition step.
In general, $\V{v}_\ell = \V{v}_0\M{P}^\ell $ gives the probability distribution after $\ell$ steps.
We indicate the state represented by $(n_0, n_1, \dots, n_d) = (M, 0, \dots, 0)$ as state zero $\ell = 0$.
Considering the system initialized to the state zero, we have that its distribution vector is $\V{v}_0 = \V{\delta}_0$ where $\V{\delta}_0 = \ (1, 0, \dots, 0)$.
Given the Markov chain described in Section~\ref{subsec:MarkovChain} with $c>0$ and $0 < p_{i \to i+1} < 1 \quad\forall i$, 
the distribution $\V{v}_k = \V{\delta}_0\M{P}^k $ converges as $k \to \infty$ to the unique stationary distribution $\V{v}_\infty$.
In fact, when $c>0$ and $p_{i \to i+1} < 1 \quad\forall i$, it is always possible to reach the state zero represented by $n_0 = M$ from any other state.
This is due to two causes: $i)$ distillation can fail, converting $n_i$ to $n_0$, for $i < d$; and $ii)$ consumption converts $n_d$ to $n_0$.
Note that in case of stalling situations arising from the fact that distillation cannot be executed (e.g., $n_1 = 1$ with $d>1$), successes on distillation can break the stall (e.g., odd number of successes on pairs with fidelity $F_0$).
Then, observe that $P_{0,0} > 0$, leading to aperiodicity of that state in the Markov chain.
By the previous consideration we have that the communicating class $\mathcal{C}$ having the state zero is closed. 
Since $\mathcal{C}$ is a communicating class it is irreducible. 
Then, by standard Markov chain theory, $\V{v}_k \to \V{v}_\infty$ as $k \to \infty$.


We can now express the stationary distribution $\V{v}_\infty$ in the more convenient $d$-dimensional form, denoted by $p_\infty(n_0,n_1,\dots,n_d)$, which gives the asymptotic probability of being in state $(n_0,n_1,\dots,n_d)$.
We define the outage probability as the probability that the stationary asymptotic state has $n_d < c$, which is the probability that is not possible to perform the consumption phase. 
More formally, the outage probability is
\begin{align}
\label{eq:outage}
    P_{out} = \sum_{\forall n_0} \sum_{\forall n_1} \cdots \sum_{n_d = 0}^{c - 1} p_\infty(n_0, n_1, \dots, n_d)\,.
\end{align}
Applying \eqref{eq:outage} within the Markov chain framework introduced in Section~\ref{subsec:MarkovChain} enables the derivation of system-level design criteria. 
As an example, for a given consumption rate $c$ and an initial state of the \ac{EPR} determined by the underlying technology, the required memory size $M$ to achieve an outage probability of $P_\mathrm{out} = 10^{-4}$ can be obtained.
This can be used not only to design quantum communication systems, but also design quantum computing modules using entanglement to perform inter-module operations like in the IBM architecture proposal~\cite{Yod25:IBM_tourgross}.
\begin{figure}[t]
    \centering
    \includegraphics[width = 0.94\columnwidth]{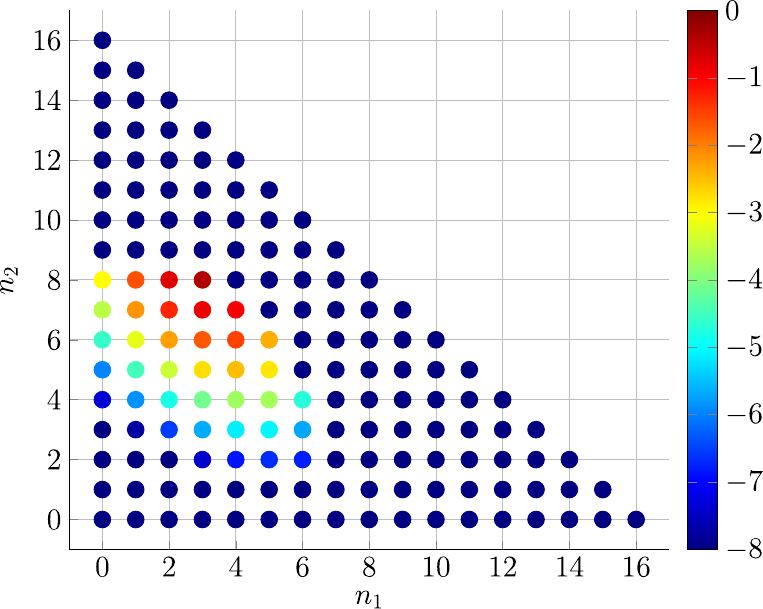}
   \caption{Asymptotic probability mass function $\log_{10} p_\infty(n_0, n_1, n_2)$ describing the \ac{EPR} memory state, with $d=2$, $F_0 = 0.9$, $c = 1$, $M = 16$. }
   \label{fig:pictExample}
\end{figure}
In Fig.~\ref{fig:pictExample}, we present a pictorial representation of the asymptotic state distribution for a system with memory $M=16$, maximum distillation steps $d=2$, consumption $c=1$, and initial raw \ac{EPR} pairs in a Werner state with fidelity $F_0 = 0.9$. 
The color scale represents $\log p_\infty(n_0, n_1, \dots, n_d)$, indicating the relative likelihood of each state in the steady regime. 
In this configuration, the system most likely converges to states where $n_2$ is between six and eight.


\subsection{Bootstrap Protocol}

In scenarios where the consumption rate $c$ is high, a more robust strategy can be adopted by introducing an initial waiting phase with no consumption before the actual consumption phase begins. This approach, referred to as the \emph{bootstrap protocol}, allows the system to accumulate resources or improve the quality of stored states prior to usage.

Let $\M{P}_c$ denote the transition probability matrix of the Markov chain described in Section~\ref{subsec:MarkovChain} under a consumption of $c$. 
The transition probability matrix corresponding to the bootstrap protocol can then be expressed as
\begin{align}
    \M{P} = \M{P}_0^{W} \M{P}_c 
\end{align}
where $W$ represents the number of waiting rounds without any consumption (i.e., the bootstrap period).
Note that $\M{P}_0$ refers to $\M{P}_c$ with $c = 0$.

This protocol enables the system to sustain the desired consumption with improved reliability, at the expense of additional latency due to the initial waiting phase. 
It is particularly advantageous when memory resources are limited, as it provides a better trade-off between reliability, latency, and memory size.

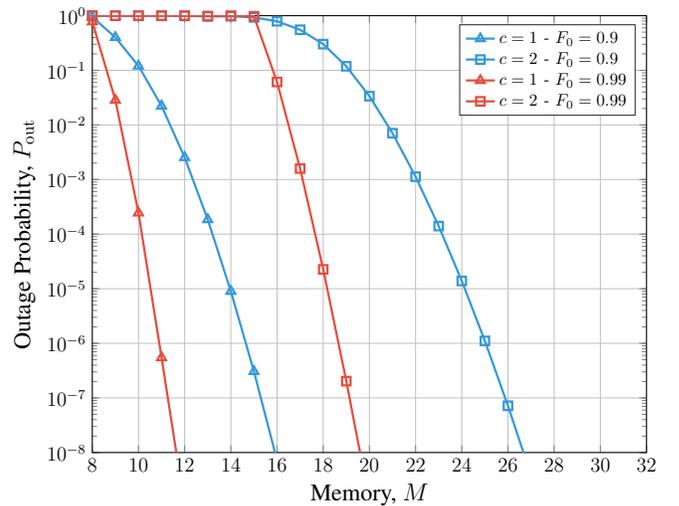
\begin{figure}[t]
    \centering
    \resizebox{\columnwidth}{!}{
%
%
\definecolor{mycolor1}{rgb}{0.00000,0.44700,0.74100}%
\definecolor{mycolor2}{rgb}{0.85000,0.32500,0.09800}%
\definecolor{mycolor3}{rgb}{0.92900,0.69400,0.12500}%
\definecolor{mycolor4}{rgb}{0.49400,0.18400,0.55600}%
\begin{tikzpicture}

\begin{axis}[%
width=4.521in,
height=3.566in,
at={(0.758in,0.481in)},
scale only axis,
xmin=8,
xmax=32,
xlabel style={font=\color{white!15!black}, font = \Large},
xlabel={Memory, $M$},
ymode=log,
ymin=1e-08,
ymax=1,
yminorticks=true,
ylabel style={font=\color{white!15!black}, font = \Large},
tick label style={black, semithick, font=\large},
ylabel={Outage Probability, $P_\mathrm{out}$},
axis background/.style={fill=white},
legend style={legend cell align=left, align=left, draw=white!15!black},
xmajorgrids=true,
ymajorgrids=true, 
]
\addplot [color=goodblue, line width=1.3pt, mark=triangle, mark size = 3.0 pt, mark options={solid, goodblue}]
  table[row sep=crcr]{%
8	0.949200955981216\\
9	0.400803849681995\\
10	0.120273347315477\\
11	0.0222593367588353\\
12	0.0025329446097517\\
13	0.000184887793431933\\
14	8.98110663409895e-06\\
15	3.05951304642669e-07\\
16	7.20270271765055e-09\\
17	1.23121665130721e-10\\
18	1.48721406779282e-12\\
19	1.3468851058269e-14\\
20	8.53795639490427e-17\\
21	4.15049538047036e-19\\
22	1.4120086989583e-21\\
23	3.73154793693312e-24\\
24	6.89199913516482e-27\\
25	9.98779216617525e-30\\
26	1.00685813613659e-32\\
27	8.01672008478083e-36\\
28	4.43433737914984e-39\\
29	1.94496770509067e-42\\
30	5.91971051714135e-46\\
31	1.43223085100677e-49\\
32	2.40185336439831e-53\\
};
\addlegendentry{$c = 1$ - $F_0 = 0.9$}

\addplot [color=goodblue, line width=1.3pt, mark=square, mark size = 2.5 pt, mark options={solid, goodblue}]
  table[row sep=crcr]{%
8	1\\
9	1\\
10	1\\
11	1\\
12	0.997944797796041\\
13	0.970013988300763\\
14	0.977901928958786\\
15	0.930127157513031\\
16	0.790563221860501\\
17	0.553060876434549\\
18	0.301632031970963\\
19	0.11849701449455\\
20	0.0336167834572757\\
21	0.00704849169546717\\
22	0.00112577698896249\\
23	0.000140268331370546\\
24	1.38800449535688e-05\\
25	1.10639232701934e-06\\
26	7.18163018537914e-08\\
27	3.82809655756353e-09\\
28	1.68768432036262e-10\\
29	6.18909168005825e-12\\
30	1.89745154020016e-13\\
31	4.88586590685788e-15\\
32	1.05766273035157e-16\\
};
\addlegendentry{$c = 2$ - $F_0 = 0.9$}

\addplot [color=goodred, line width=1.3pt, mark=triangle, mark size = 3.0 pt, mark options={solid, goodred}]
  table[row sep=crcr]{%
8	0.775232684651562\\
9	0.0284626766655105\\
10	0.000244722644850712\\
11	5.45824337050389e-07\\
12	1.04509142072503e-09\\
13	1.14010780874209e-12\\
14	3.12922915236224e-16\\
15	6.85051235932391e-20\\
16	7.79783061591269e-24\\
17	2.36223984557908e-28\\
18	5.73016164051184e-33\\
19	9.16299705052852e-38\\
20	3.74886046052222e-43\\
21	1.18810295034132e-48\\
22	1.6910284852052e-54\\
23	9.639690768081e-61\\
24	2.95869105949197e-67\\
25	6.78827980621448e-74\\
26	4.18551526053889e-81\\
27	1.94988710351362e-88\\
28	3.63041339038825e-96\\
29	3.53124254864094e-104\\
30	1.46697871709054e-112\\
31	4.71791595924217e-121\\
32	4.36143716150166e-130\\
};
\addlegendentry{$c = 1$ - $F_0 = 0.99$}

\addplot [color=goodred, line width=1.3pt, mark=square, mark size = 2.5 pt, mark options={solid, goodred}]
  table[row sep=crcr]{%
8	1\\
9	0.999999999999999\\
10	0.999999999999999\\
11	1\\
12	0.999972848368093\\
13	0.98610972004325\\
14	0.998929746067268\\
15	0.977732823406234\\
16	0.0611915187886676\\
17	0.00158745582141159\\
18	2.2618954937184e-05\\
19	2.02745998152999e-07\\
20	1.27966204409796e-09\\
21	5.80311598952726e-12\\
22	1.61055349505218e-14\\
23	2.46884519246241e-17\\
24	3.32904401279092e-20\\
25	3.91043088720221e-23\\
26	1.89271161783045e-26\\
27	1.00293361366982e-29\\
28	2.93482485004581e-33\\
29	6.73016868030693e-37\\
30	6.80114596348497e-41\\
31	1.40813044694725e-44\\
32	5.84138758036153e-49\\
};
\addlegendentry{$c = 2$ - $F_0 = 0.99$}

\end{axis}
\end{tikzpicture}%
    }
   \caption{Outage probability vs. memory size using $W=0$, and $d=2$. }
   \label{fig:Poutage}
\end{figure}

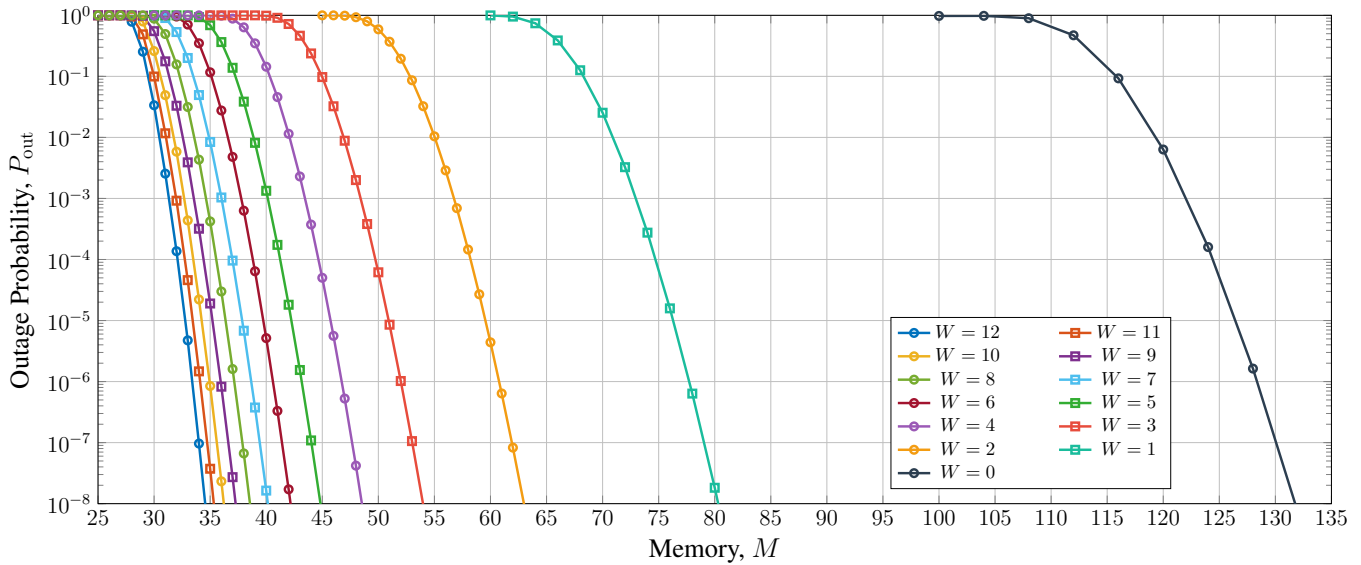
\begin{figure*}[t]
    \centering
    \resizebox{\textwidth}{!}{
%
\definecolor{mycolor1}{rgb}{0.00000,0.44700,0.74100}%
\definecolor{mycolor2}{rgb}{0.85000,0.32500,0.09800}%
\definecolor{mycolor3}{rgb}{0.92900,0.69400,0.12500}%
\definecolor{mycolor4}{rgb}{0.49400,0.18400,0.55600}%
\definecolor{mycolor5}{rgb}{0.46600,0.67400,0.18800}%
\definecolor{mycolor6}{rgb}{0.30100,0.74500,0.93300}%
\definecolor{mycolor7}{rgb}{0.63500,0.07800,0.18400}%
\begin{tikzpicture}

\begin{axis}[%
width=9in,
height=3.566in,
at={(0.758in,0.481in)},
scale only axis,
xmin=25,
xmax=135,
xlabel style={font=\color{white!15!black}, font = \Large},
xlabel={Memory, $M$},
ymode=log,
ymin=1e-08,
ymax=1,
yminorticks=true,
ylabel style={font=\color{white!15!black}, font = \Large},
tick label style={black, semithick, font=\large},
ylabel={Outage Probability, $P_\mathrm{out}$},
axis background/.style={fill=white},
legend style={
        at={(0.87,0.03)},        
        anchor=south east,       
        legend columns=2,         
        /tikz/every even column/.append style={column sep=1cm} 
    },
xmajorgrids=true,
ymajorgrids=true, 
]

\addplot [color=mycolor1, line width=1.3pt,, mark=o, mark size = 2.0 pt, mark options={solid, mycolor1}]
  table[row sep=crcr]{%
24	1.00000000000005\\
25	1.00000000000006\\
26	1.00000000000006\\
27	1.00000000000006\\
28	0.784666503303855\\
29	0.253475053047947\\
30	0.0336307472128554\\
31	0.00255028366772093\\
32	0.000136778187709508\\
33	4.73461429169119e-06\\
34	9.65785251158473e-08\\
35	1.66793801519106e-09\\
36	1.95488719496085e-11\\
37	1.84532232509277e-13\\
38	1.17038121520227e-15\\
39	6.51190000350338e-18\\
40	2.02031650500771e-20\\
41	6.16396958137539e-23\\
42	1.03172250179709e-25\\
43	1.77955323596325e-28\\
44	1.49341742300665e-31\\
45	1.48198183923658e-34\\
46	5.92437438803871e-38\\
};
\addlegendentry{$W = 12$}

\addplot [color=mycolor2, line width=1.3pt,, mark=square, mark size = 2.0 pt, mark options={solid, mycolor2}]
  table[row sep=crcr]{%
24	1.00000000000006\\
25	1.00000000000006\\
26	1.00000000000006\\
27	1.00000000000006\\
28	0.963186938086404\\
29	0.488522717336784\\
30	0.0995189662624749\\
31	0.0116847959506518\\
32	0.000919201942477674\\
33	4.59155086269779e-05\\
34	1.47072218860354e-06\\
35	3.73135175426967e-08\\
36	6.63623704311364e-10\\
37	9.24170695219527e-12\\
38	0\\
39	0\\
40	0\\
41	0\\
42	0\\
43	0\\
44	0\\
45	0\\
46	0\\
};
\addlegendentry{$W = 11$}

\addplot [color=mycolor3, line width=1.3pt,, mark=o, mark size = 2.0 pt, mark options={solid, mycolor3}]
  table[row sep=crcr]{%
24	1.00000000000006\\
25	1.00000000000006\\
26	1.00000000000006\\
27	1.00000000000006\\
28	1.00000000000006\\
29	0.791214595701571\\
30	0.259693032694384\\
31	0.0491739307842632\\
32	0.00583893382552112\\
33	0.000437147970459904\\
34	2.21505294125377e-05\\
35	8.46041412360371e-07\\
36	2.32097143696779e-08\\
37	4.89274356567318e-10\\
38	7.65769959537488e-12\\
39	0\\
40	0\\
41	0\\
42	0\\
43	0\\
44	0\\
45	0\\
46	0\\
};
\addlegendentry{$W = 10$}

\addplot [color=mycolor4, line width=1.3pt,, mark=square, mark size = 2.0 pt, mark options={solid, mycolor4}]
  table[row sep=crcr]{%
24	1.00000000000005\\
25	1.00000000000005\\
26	1.00000000000006\\
27	1.00000000000005\\
28	1.00000000000006\\
29	0.98450222868347\\
30	0.550762874884552\\
31	0.176371274462996\\
32	0.0330420345389682\\
33	0.00390456575671112\\
34	0.000319234350182901\\
35	1.90109651969574e-05\\
36	8.24557299411641e-07\\
37	2.71879361826837e-08\\
38	6.84556415096405e-10\\
39	1.35404425983537e-11\\
40	0\\
41	0\\
42	0\\
43	0\\
44	0\\
45	0\\
46	0\\
};
\addlegendentry{$W = 9$}

\addplot [color=mycolor5, line width=1.3pt,, mark=o, mark size = 2.0 pt, mark options={solid, mycolor5}]
  table[row sep=crcr]{%
24	1.00000000000005\\
25	1.00000000000005\\
26	1.00000000000004\\
27	1.00000000000005\\
28	1.00000000000004\\
29	1.00000000000004\\
30	0.881546827828711\\
31	0.493829117076566\\
32	0.156997504708574\\
33	0.0315061875874776\\
34	0.00432531890159201\\
35	0.000421075413858305\\
36	2.99074105076398e-05\\
37	1.60637091025227e-06\\
38	6.65593146794444e-08\\
39	2.14330639112079e-09\\
40	5.33429654016098e-11\\
41	0\\
42	0\\
43	0\\
44	0\\
45	0\\
46	0\\
};
\addlegendentry{$W = 8$}

\addplot [color=mycolor6, line width=1.3pt,, mark=square, mark size = 2.0 pt, mark options={solid, mycolor6}]
  table[row sep=crcr]{%
30	0.99922756568289\\
31	0.903641401191055\\
32	0.532518952468806\\
33	0.200044600421709\\
34	0.0494437191436301\\
35	0.00842228964279986\\
36	0.0010373896721765\\
37	9.5958038527649e-05\\
38	6.81328891913404e-06\\
39	3.7591582847823e-07\\
40	1.63705505527466e-08\\
41	5.7079952579508e-10\\
42	1.5925081764785e-11\\
43	0\\
44	0\\
45	0\\
46	0\\
47	0\\
48	0\\
};
\addlegendentry{$W = 7$}

\addplot [color=mycolor7, line width=1.3pt,, mark=o, mark size = 2.0 pt, mark options={solid, mycolor7}]
  table[row sep=crcr]{%
30	1\\
31	0.999999999998923\\
32	0.955108713933682\\
33	0.706208166254042\\
34	0.348633152989706\\
35	0.116635436109966\\
36	0.0276100288026295\\
37	0.00479714709093948\\
38	0.000630147739093064\\
39	6.41309351456841e-05\\
40	5.15125589102765e-06\\
41	3.30797730354557e-07\\
42	1.71631001341096e-08\\
43	7.26042377871189e-10\\
44	2.52052208510757e-11\\
45	0\\
46	0\\
47	0\\
48	0\\
};
\addlegendentry{$W = 6$}

\addplot [color=goodgreen, line width=1.3pt,, mark=square, mark size = 2.0 pt, mark options={solid, goodgreen}]
  table[row sep=crcr]{%
30	1.00000000000001\\
31	1.00000000000001\\
32	0.999999999899843\\
33	0.998103109177476\\
34	0.93342349647349\\
35	0.689224119732197\\
36	0.364759768568394\\
37	0.138224111667543\\
38	0.0385526853016371\\
39	0.00814545729388222\\
40	0.00133593825916133\\
41	0.000173535379563376\\
42	1.81460097862742e-05\\
43	1.54707030427863e-06\\
44	1.0861000318785e-07\\
45	6.32983125149052e-09\\
46	3.0838603212018e-10\\
47	1.26286256264843e-11\\
48	0\\
};
\addlegendentry{$W = 5$}

\addplot [color=HTMLlightviolet, line width=1.3pt,, mark=o, mark size = 2.0 pt, mark options={solid, HTMLlightviolet}]
  table[row sep=crcr]{%
30	1.00000000000004\\
31	1.00000000000004\\
32	1.00000000000004\\
33	0.999999999990335\\
34	0.999999993691613\\
35	0.999975378155556\\
36	0.988017807732189\\
37	0.8829379948284\\
38	0.635675643787953\\
39	0.347668689284439\\
40	0.143851907472681\\
41	0.0458028824632295\\
42	0.0114570191093036\\
43	0.002294049517167\\
44	0.000373579515062372\\
45	5.01357784256685e-05\\
46	5.60587714600414e-06\\
47	5.26986408944438e-07\\
48	4.19650499686408e-08\\
49	2.84876942838119e-09\\
50	1.65743818973404e-10\\
51	8.30292049360534e-12\\
52	0\\
53	0\\
54	0\\
};
\addlegendentry{$W = 4$}

\addplot [color=HTMLred, line width=1.3pt,, mark=square, mark size = 2.0 pt, mark options={solid, HTMLred}]
  table[row sep=crcr]{%
35	1.00000000000002\\
36	0.999999996910049\\
37	0.99999942275659\\
38	0.999999550163174\\
39	0.999672188766694\\
40	0.987575087212868\\
41	0.910902337582216\\
42	0.72130464645332\\
43	0.463608773332387\\
44	0.237593107806481\\
45	0.0974523338751386\\
46	0.0323703743966075\\
47	0.00882576746831721\\
48	0.00200053039023868\\
49	0.00038127166564104\\
50	6.16968450799647e-05\\
51	8.54804404826634e-06\\
52	1.02131859653065e-06\\
53	1.05882253965339e-07\\
54	9.57552878306946e-09\\
55	7.58900155789117e-10\\
56	5.29222868841708e-11\\
57	0\\
58	0\\
59	0\\
60	0\\
61	0\\
62	0\\
63	0\\
64	0\\
};
\addlegendentry{$W = 3$}

\addplot [color=HTMLorange, line width=1.3pt,, mark=o, mark size = 2.0 pt, mark options={solid, HTMLorange}]
  table[row sep=crcr]{%
45	0.99997058318972\\
46	0.998972419256918\\
47	0.986883147028391\\
48	0.931192371184556\\
49	0.796199826665214\\
50	0.589917474307311\\
51	0.369737017102196\\
52	0.194655579232017\\
53	0.0862985502072491\\
54	0.0324361572553766\\
55	0.0104198999330212\\
56	0.00288426905830127\\
57	0.000693178530128713\\
58	0.000145639440276839\\
59	2.69156132655072e-05\\
60	4.39934139244317e-06\\
61	6.39037955627272e-07\\
62	8.28480057968566e-08\\
63	9.62295822753377e-09\\
64	1.00480593456148e-09\\
};
\addlegendentry{$W = 2$}

\addplot [color=HTMLwatergreen, line width=1.3pt,, mark=square, mark size = 2.0 pt, mark options={solid, HTMLwatergreen}]
  table[row sep=crcr]{%
60	0.99756531812192\\
62	0.953268095960111\\
64	0.741026649716422\\
66	0.38793434802591\\
68	0.125980180103482\\
70	0.0253577513044819\\
72	0.00325424440264509\\
74	0.000275286204865804\\
76	1.5829783964533e-05\\
78	6.3539174116699e-07\\
80	1.82058798775095e-08\\
82	3.79463769627299e-10\\
84	5.84557184161164e-12\\
86	0\\
};
\addlegendentry{$W = 1$}

\addplot [color=HTMLblack, line width=1.3pt,, mark=o, mark size = 2.0 pt, mark options={solid, HTMLblack}]
  table[row sep=crcr]{%
100	0.977445471559358\\
104	0.981271566174342\\
108	0.896715146652029\\
112	0.470976864817445\\
116	0.0927889646439182\\
120	0.00633433228129578\\
124	0.000159965337631153\\
128	1.6318501069277e-06\\
132	7.2794321973638e-09\\
136	1.5168279559305e-11\\
140	0\\
};
\addlegendentry{$W = 0$}

\end{axis}
\end{tikzpicture}%
    }
   \caption{Outage probability vs. quantum memory size using the bootstrap protocol with parameter $W$. Parameters: $c=13$, $F_0 = 0.9$, and $d=2$. }
   \label{fig:PoutSurface}
\end{figure*}


\section{Numerical Results}\label{sec:NumRes}

In this section we present some results to give an idea of what can be done with the proposed framework.

\paragraph{Quantum Memory Dimensioning}
In Fig.~\ref{fig:Poutage}, we illustrate the outage probability $P_\mathrm{out}$ for two entanglement distribution schemes: one with an initial fidelity of $F_0 = 0.9$ and another with $F_0 = 0.99$. 
The figure shows how the outage probability decreases as the memory size increases, for two different values of the consumption rate $c$. 
Given a target outage probability, this analysis enables the design of the memory size required to meet specific reliability constraints.
For instance, if the target outage probability is $P_\mathrm{out} = 10^{-3}$, a memory size of $M = 10$ qubits is required for the entanglement distribution scheme with $F_0 = 0.99$, while $M = 13$ qubits are needed for the scheme with $F_0 = 0.9$, assuming a consumption of $c = 1$ entangled qubits.


\paragraph{Trade-off between memory and latency}
As another illustrative example, we may require $c = n$ entangled pairs when operating on a codeword of an $n$-qubits quantum error-correcting code.
Specifically, consider the case of a $[[13, 1, 3]]$ surface code \cite{BraKit:98, ForValChi24:JSAC}, which in our setting results in the consumption of $c = 13$ entangled pairs.
Fig.~\ref{fig:PoutSurface} illustrates the resulting trade-off between the memory size $M$ and the latency $W$ introduced by the bootstrap protocol, using $F_0 = 0.9$ and $d=2$.
From this plot we can observe that for a requirement of $P_\mathrm{out} = 10^{-3}$, a system without bootstrap ($W=0$) requires a memory of at least $M=123$ qubits, while waiting $W=3$ rounds before consumption requires a memory of at least $M = 49$ qubits, waiting $W=6$ rounds requires at least $M = 38$, and waiting $W=12$ requires at least $M=32$.
This analysis provides a framework for optimizing quantum memory design with respect to the trade-off between storage capacity and operational delay, i.e., how many communication rounds can elapse before the stored entangled qubits must be consumed.
Such a trade-off is not only relevant for quantum communication systems, where memory coherence and throughput are critical, but also for the dimensioning of modular quantum computing architectures.
As an example, this approach could offer guidance in the design of quantum distributed computing systems that employ quantum error correction and rely on entanglement-based inter-module operations~\cite{Yod25:IBM_tourgross}.




\section{Conclusions}\label{sec:conclusions}
In this work, we proposed a framework for dimensioning quantum memories capable of storing and managing high-fidelity entangled pairs for use in systems operating with quantum error correcting codes. 
We modeled the stochastic evolution of stored \ac{EPR} pairs when \ac{EPR} packets are available in each communication round using a Markov chain approach.
In this way, we established a quantitative link between memory capacity, initial entanglement fidelity, maximum number of distillation steps, and other system parameters.
Through numerical analysis, we showed how the outage probability and storage requirements vary with fidelity, consumption rate, and operational latency, providing practical guidelines for optimizing memory size. 
As a future direction, we aim to include decoherence, also tailoring the scheme to counteract this phenomena.


\section*{Acknowledgment}

This work was partially supported by the European Union under the Italian National Recovery and Resilience Plan (NRRP) of NextGenerationEU, HPC National Centre for HPC, Big Data and Quantum Computing (CN00000013).

\bibliographystyle{IEEEtran}
\bibliography{Files/IEEEabrv,Files/StringDefinitions,Files/StringDefinitions2,Files/refs}

@STRING{IEEE_J_JSAC       = "{IEEE} J. Sel. Areas Commun."}

@STRING{IEEE = {The Institute of Electrical and Electronics Engineers}}

@article{BraKit:98,
	author = {Bravyi, S. B. and Kitaev, A. Yu.},
	title = {Quantum codes on a lattice with boundary},
    eprint={quant-ph/9811052},
      archivePrefix={arXiv},
      primaryClass={quant-ph},
journal={arXiv preprint quant-ph/9811052},
      doi = {https://doi.org/10.48550/arXiv.quant-ph/9811052},
	year = {1998}}

@ARTICLE{ForValChi24:JSAC,
  author={Forlivesi, Diego and Valentini, Lorenzo and Chiani, Marco},
  journal=IEEE_J_JSAC, 
  title={Logical Error Rates of {XZZX} and Rotated Quantum Surface Codes}, 
  year={2024},
  volume={42},
  number={7},
  pages={1808-1817},
  doi={10.1109/JSAC.2024.3380088}
}

@article{Yod25:IBM_tourgross,
    title={Tour de gross: A modular quantum computer based on bivariate bicycle codes}, 
    author={Theodore J. Yoder and Eddie Schoute and Patrick Rall and others},
    year={2025},
    eprint={2506.03094},
    archivePrefix={arXiv},
    primaryClass={quant-ph},
    journal={arXiv preprint arXiv:2506.03094},
}

@misc{rfc9340,
    series =    {Request for Comments},
    number =    9340,
    howpublished =  {RFC 9340},
    publisher = {RFC Editor},
    doi =       {10.17487/RFC9340},
    author =    {Wojciech Kozlowski and Stephanie Wehner and Rodney Van Meter and others},
    title =     {{Architectural Principles for a Quantum Internet}},
    pagetotal = 37,
    year =      2023,
    month =     mar,
}

@article{Ben93:teleporting,
  title={Teleporting an unknown quantum state via dual classical and {E}instein-{P}odolsky-{R}osen channels},
  author={Bennett, Charles H and Brassard, Gilles and Cr{\'e}peau, Claude and others},
  journal={Physical review letters},
  volume={70},
  number={13},
  year={1993},
  publisher={APS}
}

@article{Wer89:WernerState,
  title={Quantum states with {E}instein-{P}odolsky-{R}osen correlations admitting a hidden-variable model},
  author={Werner, Reinhard F},
  journal={Physical Review A},
  volume={40},
  number={8},
  pages={4277},
  year={1989},
  publisher={APS}
}

@article{Ben96:purification,
  title={Purification of noisy entanglement and faithful teleportation via noisy channels},
  author={Bennett, Charles H and Brassard, Gilles and Popescu, Sandu and others},
  journal={Physical review letters},
  volume={76},
  number={5},
  pages={722},
  year={1996},
  publisher={APS}
}

@article{Zha02:WernerPrep,
  title={Experimental preparation of the Werner state via spontaneous parametric down-conversion},
  author={Zhang, Yong-Sheng and Huang, Yun-Feng and Li, Chuan-Feng and Guo, Guang-Can},
  journal={Physical Review A},
  volume={66},
  number={6},
  pages={062315},
  year={2002},
  publisher={APS}
}

@article{Deu96:purification,
  title={Quantum privacy amplification and the security of quantum cryptography over noisy channels},
  author={Deutsch, David and Ekert, Artur and Jozsa, Richard and others},
  journal={Physical review letters},
  volume={77},
  number={13},
  pages={2818},
  year={1996},
  publisher={APS}
}

@article{Cisco25:DataCenter,
      title={Quantum Data Center Infrastructures: A Scalable Architectural Design Perspective}, 
      author={Hassan Shapourian and Eneet Kaur and Troy Sewell and others},
      year={2025},
      eprint={2501.05598},
      archivePrefix={arXiv},
      primaryClass={quant-ph},
      journal={arXiv preprint arXiv:2501.05598}
}

@article{Yoder25:EPRdiversiModuli,
      title={Universal adapters between quantum {LDPC} codes}, 
      author={Esha Swaroop and Tomas Jochym-O'Connor and Theodore J. Yoder},
      year={2025},
      eprint={2410.03628},
      archivePrefix={arXiv},
      primaryClass={quant-ph},
    journal={arXiv preprint arXiv:2410.03628}
}

@article{WehElkHan18:QInternet,
	author = {Wehner, Stephanie and Elkouss, David and Hanson, Ronald},
	journal = {Science},
	number = {6412},
	title = {Quantum internet: A vision for the road ahead},
	volume = {362},
	year = {2018}
}

@article{Pom22:experimentalQI,
  title={Experimental demonstration of entanglement delivery using a quantum network stack},
  author={Pompili, Matteo and Delle Donne, Carlo and te Raa, Ingmar and others},
  journal={npj Quantum Information},
  volume={8},
  number={1},
  pages={121},
  year={2022},
  publisher={Nature Publishing Group UK London}
}

@article{Cac19:QInternet,
  title={Quantum internet: networking challenges in distributed quantum computing},
  author={Cacciapuoti, Angela Sara and Caleffi, Marcello and Tafuri, Francesco and Cataliotti, Francesco Saverio and Gherardini, Stefano and Bianchi, Giuseppe},
  journal={IEEE Network},
  volume={34},
  number={1},
  pages={137--143},
  year={2019},
  publisher={IEEE}
}

@ARTICLE{Mun2015:QRepeater,
  author={Munro, William J. and Azuma, Koji and Tamaki, Kiyoshi and Nemoto, Kae},
  journal={IEEE Journal of Selected Topics in Quantum Electronics}, 
  title={Inside Quantum Repeaters}, 
  year={2015},
  volume={21},
  number={3},
  pages={78-90},
  doi={10.1109/JSTQE.2015.2392076}
}

@inproceedings{els24:fidelity,
  title={On the Fidelity Distribution of Purified Link-level Entanglements},
  author={Elsayed, Karim S and KhudaBukhsh, Wasiur R and Rizk, Amr},
  booktitle={IEEE Int. Conf. on Commun.},
  pages={485--490},
  year={2024}
}

@article{dav24:entanglement,
  title={Entanglement buffering with two quantum memories},
  author={Davies, Bethany and I{\~n}esta, {\'A}lvaro G and Wehner, Stephanie},
  journal={Quantum},
  volume={8},
  pages={1458},
  year={2024},
  publisher={Verein zur F{\"o}rderung des Open Access Publizierens in den Quantenwissenschaften}
}

@article{Ter:15,
	author = {Terhal, Barbara M.},
	journal = {Rev. Mod. Phys.},
	month = {Apr},
	pages = {307--346},
	title = {Quantum error correction for quantum memories},
	volume = {87},
	year = {2015},
    issue = {2},
    publisher = {American Physical Society},
    doi = {10.1103/RevModPhys.87.307}
}

@article{Bra24:BBCodes,
  author       = {Bravyi, Sergey and Cross, Andrew W. and Gambetta, Jay M. and others},
  title        = {High-threshold and low-overhead fault-tolerant quantum memory},
  journal      = {Nature},
  volume       = {627},
  number       = {8005},
  pages        = {778--782},
  year         = {2024},
  doi          = {10.1038/s41586-024-07107-7},
  publisher    = {Nature Publishing Group}
}

@ARTICLE{ValForChi25:CylMob,
  author={Valentini, Lorenzo and Forlivesi, Diego and Chiani, Marco},
  journal={IEEE Trans. Inf. Theory.}, 
  title={Cylindrical and {M\"obius} Quantum Codes for Asymmetric {Pauli} Errors}, 
  year={2025},
  volume={71},
  number={5},
  pages={3766-3778},
  doi={10.1109/TIT.2025.3546769}
}


\end{document}